# Experimental study on the influence of binders on plastic deformation during sintering of cubic boron nitride powder under high pressure and high temperature

Su-Gon Kim, Gwang-Il Jon, Song-Jin Im

Faculty of Physics, **Kim Il Sung** University, Daesong District, Pyongyang, DPR Korea

**Abstract**

In this work, sintered polycrystalline cubic boron nitride (cBN) compacts, with titanium carbonitride ($TiC_{0.7}N_{0.3}$) and titanium nitride (TiN), respectively, as binders, were prepared at temperature of 1450°C and pressure of 5.5 GPa during 3 minutes, and the influence of binders on the plastic deformation of cBN powder due to the sintering process was experimentally investigated by X-ray diffraction (XRD) analysis. It was shown that the sintered compact with titanium carbonitride as a binder displays more intense plastic deformation of cBN grains than that with titanium nitride binder. This result indicated that if the binders are different, then products formed during sintering of cBN powders under high pressure and high temperature (HP-HT) are different so that the stress concentration at the cBN grain boundaries and a pinning effect, reducing the mobility of the dislocations and preventing annealing, are different in two cases.

## 1. Introduction

cBN compacts have outstanding properties such as high oxidation resistance, high thermal stability, chemical inactivity for iron and iron group alloys, so it is widely used for machining of hard materials. In recently years, therefore, study for synthesizing of cBN compact is progressed in high key [1-4].

During the sintering process under HP-HT, not only new products are produced by chemical reactions between cBN and binder phases, but also the cBN powder particles are subjected to the plastic deformation. These phenomena have great influence on properties of sintered cBN compact.

Casanova et al. [5] studied the kinetics of the plastic deformation of cBN occurred during the sintering process of cBN powder with and without a binder under high pressure of 8GPa and high temperature of 1850°C for different time spans by investigating the width of the (331) XRD peak and Raman peak, and consequently showed that the time span for the maximum width of the (331) XRD peak while sintered with Al binder is longer than one while without a binder.

In our present work, from the fact that if cBN crystals are deformed plastically, the widths of the (220) XRD peak and (331) XRD peak are remarkably broadened[5], the effect of kinds of binder on the plastic deformation of cBN crystal particles occurred during the sintering process of cBN powder under HP-HT was studied.

## 2. Experimental details

### 2.1. Starting materials

Starting materials and their composition are given in table 1.

**Table 1. Grain size and sample composition of starting materials**

| Starting material. | | cBN powder | TiN powder | Ti($C_{0.7}N_{0.3}$) powder | Al powder |
|---|---|---|---|---|---|
| average grain size (μm) | | <1 | <1 | <1 | <1 |
| mixing ratio (vol %) | № 1 | 70 | | 18 | 12 |
| | № 2 | 70 | 18 | | 12 |

The binder powders were weighed and mixed into cBN powders in the mixing ratio as shown in Table 1. The starting powders were mixed with ethanol for 1 h in an agate mortar. Mixing was carried out with care to avoid contamination from the agate mortar. The dried powders were sieved through a 500 μm sieve. Green compact disks of 10mm in diameter and 3mm in thickness were formed with a compacting pressure of 300 MPa.

The cemented carbide (WC-8wt. %Co) compact disks with 10mm diameter and 1.5mm thickness were used as substrates.

The green compact disks and cemented carbide disks were surrounded with 0.15 mm thick Ni foils as shown in Fig. 1. This sample assembly was submitted to thermal treatment under vacuum of $10^{-3}$ Torr and temperature of 600 ℃ for 1 h to deaerate.

### 2.2. High pressure sintering

The high pressure sintering experiment was conducted in a special hydraulic press, using the toroidal type hard-metal anvils with concavity of diameter of 31mm and depth of 6.5mm as high pressure apparatus [6]. The scheme of the reaction cell for high pressure sintering is shown in Fig. 2. Hexagonal boron nitride (hBN) was used to insulate between the graphite heater and sample assembly.

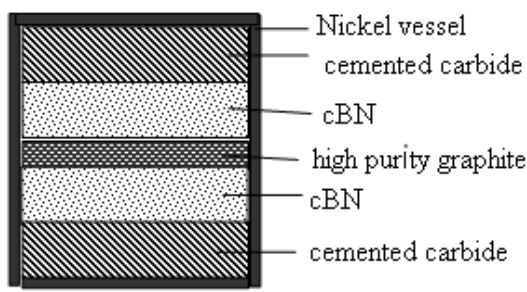

Fig. 1. Sample assembly

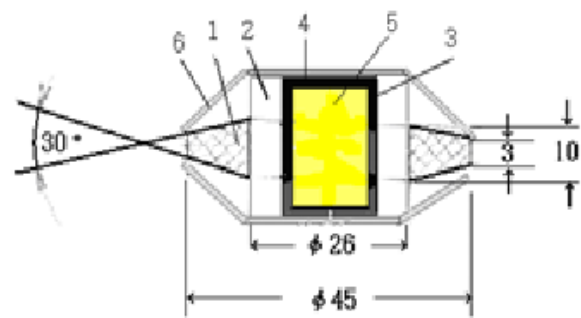

**Fig. 2. Scheme of the reaction cell for high pressure sintering**
1-gasket, 2-pressure medium, 3-graphite heater,
4-graphite cap, 5-sample assembly, 6-iron cap

The samples were loaded up to 5GPa at room temperature, then heated up to 1700K and kept in such an environment for 3 minutes. After that, temperature was lowered and then pressure was released. The pressure was calibrated using the standard bismuth, thallium and barium phase transition procedure. The temperature was generated by indirect heating in a graphite heater furnace and the calibration was performed using a Pt/Pt–13%Rh thermocouple.

After sintering, the compacts were approximately 3 mm thick with a diameter of about 9 mm.

## 3. Results and discussion

XRD was used to investigate the plastic deformation of the grains based on the broadening of the (220) line and (331) line and, also, to analyze the existence of different crystalline phases formed during sintering. The XRD measurements were carried out in Smart Lab diffractometer, using CuKα radiation in the angular region between 20° and 150°.

### 3.1. XRD analysis

Fig.3 and 4 show the XRD patterns taken from compacts of cBN-Ti($C_{0.7}N_{0.3}$)-Al composite and cBN-TiN-Al composite under HP-HT. XRD pattern for cBN-Ti($C_{0.7}N_{0.3}$)-Al composite indicates that all of Al react on the other materials, and new products such as AlN, $AlB_2$, $TiB_2$, TiN, TiC, and $B_4C$ are formed, besides cBN and Ti($C_{0.7}N_{0.3}$), during sintering of cBN-Ti($C_{0.7}N_{0.3}$)-Al composite under pressure of 5GPa and temperature of 1700K. Those new products are hard materials and may form a frame strongly combining with cBN particles. According to the results of quantitative analysis by using Jade, X-ray analysis program, the total amounts of cBN, Ti($C_{0.7}N_{0.3}$), TiN, AlN, TiC, $AlB_2$, $TiB_2$ and $B_4C$ are 63.4 vol%, 14.6 vol%, 0.5 vol%, 13.2 vol%, 1.5 vol%, 2.3 vol%, 2.2 vol% and 2.2 vol%, respectively.

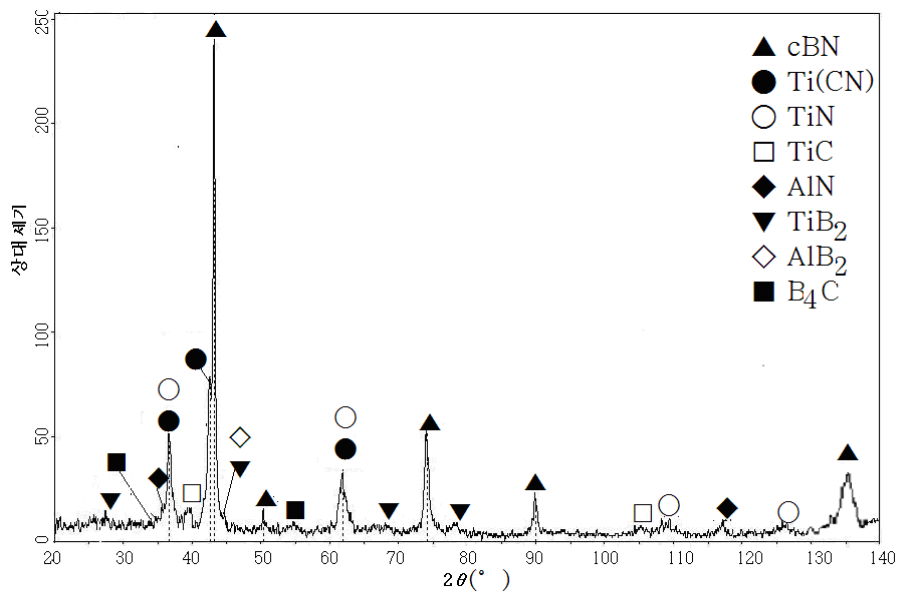

**Fig. 3. XRD patterns of cBN-Ti($C_{0.7}N_{0.3}$)-Al composite sintered at 5GPa, 1700K.**

Also, XRD pattern for cBN-TiN-Al composite indicates that all of Al react on the other materials, and besides cBN and TiN, new products such as AlN, $AlB_2$ and $TiB_2$ are formed during sintering of cBN-Ti($C_{0.7}N_{0.3}$)-Al composite under pressure of 5GPa and temperature of 1700K. Those new products are also hard materials and may form a frame strongly combining with cBN particles. According to quantitative analysis results by using Jade, X-ray analysis program, the total amounts of cBN, TiN, TiN, AlN, $AlB_2$ and $TiB_2$ are 64.3 vol%, 15.9 vol%, 12.7 vol%, 3.2 vol% and 2.9 vol%, respectively.

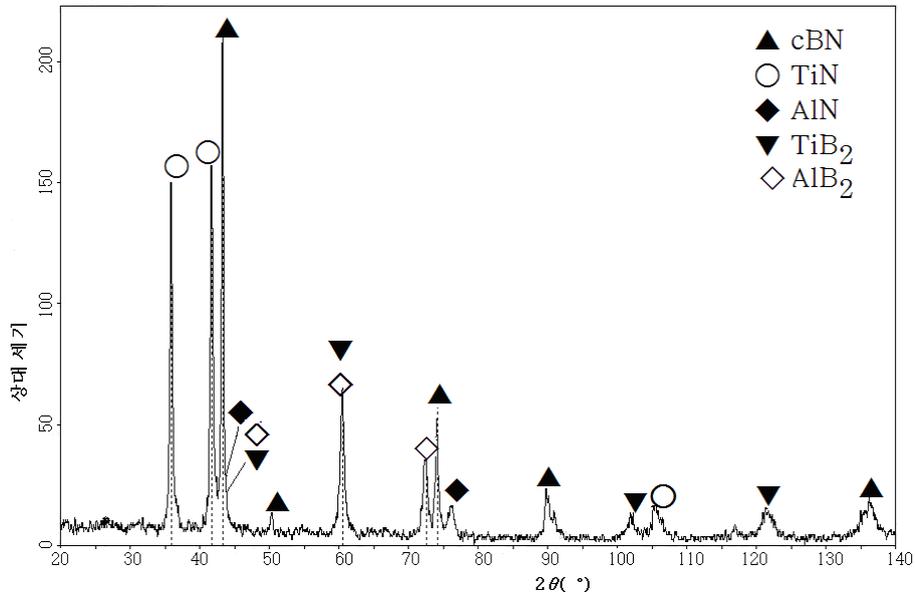

**Fig. 4. XRD patterns of cBN-TiN-Al composite sintered at 5GPa, 1700K.**

### 3.2. The characteristics of the plastic deformation of cBN grains

Fig. 5 shows the diffraction patterns of samples in the regions of the (220) line and the (331) line of cBN. Table 2 shows the full widths at half-maximum (FWHM) of the (220) peak and the (331) peak.

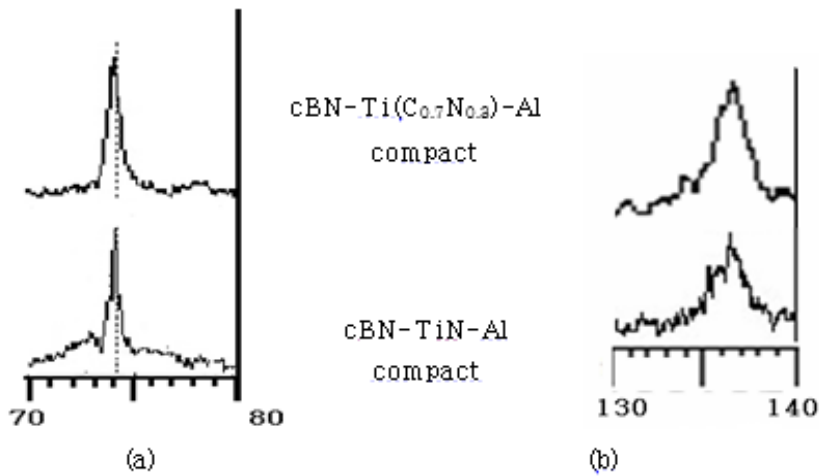

**Fig. 5. XRD patterns of (220) line (a) and (331) line (b)**

Table 2. Results of measurement of FWHMs of the (220) peak and (331) peak

| samples | cBN-Ti($C_{0.7}N_{0.3}$)-Al compact | cBN-TiN-Al compact |
|---|---|---|
| FWHM of (220) line | 0.563 | 0.363 |
| FWHM of (331) line | 0.933 | 0.620 |

As can be seen from Fig. 5 and table 2, FWHM of the (331) peak for cBN-Ti($C_{0.7}N_{0.3}$)-Al compact is larger than that for cBN-TiN-Al compact. This, probably, is due to the difference of the binders. In fact, according to XRD analysis, during HP-HT sintering, Al, Ti and C of cBN-Ti($C_{0.7}N_{0.3}$)-Al compact reacts with N and B of cBN to form refractory compounds AlN, $AlB_2$, $TiB_2$, TiN, TiC and $B_4C$, and, Al and Ti of cBN-TiN-Al compact reacts with N and B of cBN to form AlN, $AlB_2$ and $TiB_2$. Moreover, the chemical reaction with cBN grains is most likely to occur at the defective parts of the cBN grain surface, especially at dislocations. This can produce a pinning effect, reducing the mobility of the dislocations. In two cases the difference of the refractory compounds formed at cBN grain boundaries causes the difference of the stress concentration at cBN grain boundaries. In the first case the plastic deformation of cBN grains is probably larger than that in the second case because the more B and N atoms take part in reaction at the cBN grain boundaries and the stress concentration is greater. Therefore, it is preferable to choose the binders which cause the greater plastic deformation of cBN grains.

## 4. Conclusion

In this work, sintered polycrystalline cBN compacts, with titanium carbonitride and titanium nitride, respectively, as binders, were prepared at temperature of 1450°C and pressure of 5.5 GPa during 3 minutes, and the influence of binders on the plastic deformation of cBN powder due to the sintering process was experimentally investigated by XRD analysis. It was shown that the sintered compact with titanium carbonitride as a binder displays more intense plastic deformation of cBN grains than that with titanium nitride binder. This, probably, is because, in the case of using Ti($C_{0.7}N_{0.3}$) as a binder, more and harder products are formed during the sintering process under HP-HT than that in the case of using TiN as a binder, so that the mobility of the dislocations in cBN grains is reduced and the stress concentration at the cBN grain boundaries is increased.